\documentclass[11pt]{article}

\usepackage{a4}
\usepackage{amsmath}
\usepackage{amsfonts}
\usepackage{latexsym}
\usepackage{epsfig}
\begin{document}
\baselineskip 0.6cm

\begin{center}
{\Large \bf 5D $SU(3)_W$ unification at TeV and cancellation of local gauge
anomalies with split multiplets
\footnote{Talk at Summer Institute 2002, Fuji-Yoshida, Japan, 
13-20 August 2002.} }

\vskip 0.5cm

{\large \bf  Hyun Min Lee\footnote{E-mail: minlee@th.physik.uni-bonn.de} }

{\it Physikalisches Institut der Universit${\ddot a}$t Bonn,
Nussalle 12, D53115 Bonn, Germany}

\vskip 0.5cm

\abstract{
We consider the 5D gauge unification of $SU(2)_L\times U(1)_Y$ into
$SU(3)_W$ at a TeV scale. Compactification of the extra dimension
on an orbifold $S^1/(Z_2\times Z'_2)$ allows fixed points
where $SU(2)_L\times U(1)_Y$ representations can be assigned.
We explain the long proton lifetime
and the top-bottom mass hierarchy geometrically.
We also show that local gauge anomalies on the orbifold can be exactly
cancelled by a 5D Chern-Simons term with a jumping coefficient.
}

\end{center}

\noindent
We know that the gauge couplings of the Standard Model(SM) undergo
the logarithmical runnings to high energy scales
due to quantum corrections.
In the Minimal Supersymmetric Standard Model(MSSM), for instance,
the gauge couplings become unified at $M_{GUT}=2\times 10^{16}$ GeV within the
experimental error bound\cite{gut}.
The large hierarchy of scales is needed to get a large $\alpha_s$ and
${\rm sin}^2\theta_W\simeq 0.231$
at $M_Z$ with the unification of gauge couplings 
at $M_{GUT}$\footnote{In the case with SU(5) unification,
note that $\alpha_{1,2,3}\simeq \frac{1}{25}$ and
${\rm sin}^2\theta_W=\frac{3}{8}=0.375$.}. In the MSSM, the so called gauge
hierarchy is maintained by softly broken supersymmetry against radiative 
corrections to the Higgs mass.

To alleviate the gauge hierarchy problem in the GUTs with a large gauge group,  
we can consider the low energy unification of $SU(2)_L\times U(1)_Y$ into
$SU(3)_W$\cite{weinberg}. 
Then, the lepton sector($L,e^c$) and the Higgs sector
are successfully embedded into $\bf 3$'s
of $SU(3)_W$ with the hypercharge operator $Y={\rm diag.}(-1/2,-1/2,+1)$.
Moreover, we have ${\rm sin}^2\theta_W=0.25$ at the unification
scale, which is so close to its experimental value at $M_Z$ 
and thus does not need large
logarithms to run. However, it is not possible to accommodate quark fields
without an extra $U(1)$ gauge group beyond $SU(3)_W$ due to their fractional
hypercharges in units of $\frac{1}{2}$.

There has been recently a lot of attention to the GUT models 
on orbifolds with extra dimensions\cite{kawamura,kkl1}.
The main virtue of the GUT orbifolds is that the breaking of gauge symmetry 
and/or supersymmetry
and the doublet-triplet splitting can be performed at the same time
by the geometrical boundary conditions. 
Along this line, the 5D $SU(3)_W$ gauge theory has been considered 
on an orbifold $S^1/(Z_2\times Z'_2)$ with its radius $R$\cite{su3,hall,kkl2}.
If we impose on the bulk gauge fields different charges 
under the two $Z_2$ reflection
symmetries, then the bulk gauge symmetry can be broken down
to the electroweak gauge symmetry at the TeV-sized compactification scale.
The corresponding boundary conditions of the gauge fields 
$A_M=(A_\mu,A_5)(\mu=0,1,2,3)$ are written in terms of one relection $Z_2$
and one twist $T=Z_2\times Z'_2$ as
\begin{eqnarray}
A_\mu(y)&=&Z_2 A_\mu(-y) Z_2^{-1}=TA_\mu(y+\pi R)T^{-1}, \\
A_5(y)&=&-Z_2 A_5(-y) Z_2^{-1}=TA_5(y+\pi R)T^{-1}
\end{eqnarray}
where $Z_2={\rm diag.}(1,1,1)$ and $T={\rm diag.}(1,1,-1)$. 
The 5D fundamental scale($M_s$), which is regarded as the unification scale, 
is also chosen to be not far above the TeV scale
such that the logarithmic Kaluza-Klein
corrections to the running of gauge couplings can predict
a correct ${\rm sin}^2\theta_W$ at low energies.
Thus, there is no big desert between the unification scale and the weak scale
but we still have to cope with the rapid proton decay
in generic models with such a low fundamental scale.

After the orbifold compactification, there appear two fixed points
where the local gauge symmetry is different. Only the $SU(2)_L\times U(1)_Y$
gauge symmetry remains at one of fixed points($y=\pi R/2$),
which is denoted as $A$, where
quark fields can be located. On the other hand, the full bulk gauge symmetry
is respected at the other fixed point($y=0$), which is denoted as $O$.
Therefore, the location of the lepton sector
or the Higgs sector in this model, is not determined
from a field theoretic point of view.
In this paper, to avoid the rapid proton decay,
we take the minimal embedding
by splitting leptons and quarks maximally in the extra dimension:
leptons at $O$ and quarks at $A$\cite{su3,hall,kkl2}.
Furthermore, to give the top-bottom mass hierarchy,
we take the asymmetric embedding for the Higgs
sector: a up-type Higgs($H_u$) at $A$
and a down-type Higgs($H_d$) in the bulk\cite{kkl1,kkl2}. Here we note that 
$H_d$ should come from a bulk $\bf {\bar 6}_H$ to give reasonable lepton masses
at $O$. 

With the embedding of the SM particles on the orbifold, 
we consider the running of gauge couplings above the
compactification scale($M_c=R^{-1}$) due to KK modes.
The electroweak gauge symmetry
allows arbitrary brane gauge kinetic terms at $A$, which could spoil the low
energy prediction of this orbifold model. But, with the assumption of strong
coupling constants at the compactification scale, i.e. $M_s R= {\cal O}(100)$,
we can make a prediction on the Weinberg angle at $M_Z$ with KK corrections
above $M_c$\cite{hall,kkl2}:
\begin{eqnarray}
{\rm sin}^2\theta_W(M_Z)=0.25-\frac{3}{8\pi}\alpha_{em}
\bigg[{\tilde B}\ln\frac{M_s}{M_c}+B\ln\frac{M_c}{M_Z}\bigg]
\end{eqnarray}
where $B=b_g-b_{g'}/3$ and ${\tilde B}={\tilde b}_g-{\tilde b}_{g'}/3$.
Here $b$'s denote the beta function coefficients for the SM zero modes 
and $\tilde b$'s denote those for the KK modes.
As a result, in the non-SUSY case, the fundamental scale
becomes $70-80$ TeV, depending on one or two Higgs fields.
In the MSSM case, we
obtain $M_s=1.9-3.4\times 10^4$ TeV, depending on asymmetric or symmetric
embedding of the Higgs sector.

Now let us look into the local gauge anomalies in our model
for consistency. Since we have leptons and quarks located at different fixed
points, it seems that there could exist corresponding local gauge anomalies
at each fixed point. Moreover, the gauge anomalies coming
from a single bulk Higgsino field are equally splitted at both fixed
points\cite{anomaly},
so that there could be the remaining local gauge anomalies even after a brane
Higgsino field with opposite charge is taken into account.
In any case, it turns out that the total remaining local gauge anomalies
are cancelled exactly by the variation of a bulk non-abelian Chern-Simons
term with a proper normalization $c=3c_l+c_H=7$:
\begin{eqnarray}
{\cal L}_{CS}=\frac{c}{128\pi^2}\epsilon(y){\rm tr}\bigg(AF^2-\frac{1}{2}A^3 F
+\frac{1}{10}A^5\bigg)
\end{eqnarray}
where $\epsilon(y)$ is the sign function with periodicity $\pi R$.
This Chern-Simons term is parity odd and thus explicitly breaks the
$Z_2$ parity symmetries, which implies that the gauge symmetry on the orbifold
is maintained at the quantum level only at the price of the parity violation.
It is also shown that gravitational mixed anomalies of $U(1)_Y$, which could
appear only at $A$, cancel between quarks and Higgsino fields separately
without the aid of a bulk Chern-Simons term\cite{kkl2}. 

To upshot, it is shown that in the 5D $SU(3)_W$ unification model compactified
on $S^1/(Z_2\times Z'_2)$, the quark
sector is also accommodated at the fixed point retaining only the electroweak
gauge symmetry. We also considered SM multiplets split in other locations 
of the extra dimensions to explain the proton stability and the $t-b$ mass 
hierarchy. This splitting still gives rise to a consistent gauge theory with
the introduction of a 5D Chern-Simons term. Our model also helps to
predict ${\rm sin}^2\theta_W$ at low energies through KK modes without the need
of a large hierarchy.   

\vskip 0.5cm
\noindent
{\it Acknowledgements}: I would like to thank the organizers of the Summer 
Institute 2002 in Japan for their hospitality. This work is supported 
by the BK21 program of Ministry of Education in Korea.


\end{document}